\documentclass[a4paper,11pt]{article}
\usepackage{pos}

\title{Neutrino mass ordering obfuscated by the NSI}

\author[a]{Francesco Capozzi}
\author*[b]{Sabya Sachi Chatterjee}
\author[c,d]{Antonio Palazzo}
\affiliation[a]{Max-Planck-Institut f\"ur Physik (Werner-Heisenberg-Institut), F\"ohringer Ring 6, 80805 M\"unchen, Germany}
\affiliation[b]{Institute for Particle Physics Phenomenology, Department of Physics, Durham University, DH1 3LE, UK}
\affiliation[c]{Dipartimento Interateneo di Fisica``Michelangelo Merlin," Via Amendola 173, 70126 Bari, Italy}
\affiliation[d]{Istituto Nazionale di Fisica Nucleare, Sezione di Bari, Via Orabona 4, 70126 Bari, Italy}

\emailAdd{capozzi@mpp.mpg.de}
\emailAdd{sabya.s.chatterjee@durham.ac.uk}
\emailAdd{palazzo@ba.infn.it}
\abstract{The current data of the two long-baseline experiments T2K and NO$\nu$A indicate a $2.4\sigma$ preference of normal neutrino mass ordering over the inverted mass ordering in case of standard $3\nu$ analysis. However such a claim does not remain unaltered in presence of neutrino non-standard interactions (NSI) involving $e-\tau$ sector. We find that the sensitivity towards the neutrino mass ordering gets completely lost. So a rubust claim of the neutrino mass ordering requires a careful consideration of the impact of NSI.}

\FullConference{%
  40th International Conference on High Energy physics - ICHEP2020\\
  July 28 - August 6, 2020\\
  Prague, Czech Republic (virtual meeting)
}

\begin{document}
\maketitle

\section{Introduction}
\vspace*{-.4cm}
\noindent Determination of the neutrino mass ordering (NMO) is one of the biggest priorities in the intensity frontier of high energy particle physics. To accomplish that goal a lot of efforts are being put together with the atmospheric, solar, reactor, and accelerator neutrinos.
In the standard 3-flavor framework, NMO is defined to be normal if $m_1<m_2<m_3$, and inverted if $m_3<m_1<m_2$, where $m_1$, $m_2$, and $m_3$ are the masses of the three neutrino mass eigenstates $\nu_1$, $\nu_2$, and $\nu_3$ respectively. Interestingly, two long-baseline experiments T2K and NO$\nu$A are playing a leading role in this direction and provide a $\sim2.4\sigma$ indication in favor of normal ordering (NO) which we find in this work.
In addition, we examine how the situation looks like in presence of non-standard interactions (NSI) of neutrinos with a special focus on the non-diagonal flavor changing type $\varepsilon_{e\tau}$ and $\varepsilon_{e\mu}$. We find that the present indication of NO in the standard 3-flavor framework gets completely vanished in the presence of NSI of the flavor changing type involving the $e-\tau$ flavors.
\section{Theoretical framework}
\vspace*{-.4cm}
\noindent NSI may represent the low-energy manifestation of high-energy physics involving new
heavy states (for a review see~\cite{Farzan:2017xzy,Dev:2019anc}) or, 
alternatively, they can be mediated via the light mediators~\cite{Farzan:2015doa,Farzan:2015hkd}.
As first recognized in~\cite{Wolfenstein:1977ue}, NSI of type neutral current (NC) can modify the dynamics~\cite{Wolfenstein:1977ue,Mikheev:1986gs} of the neutrino flavor conversion in matter. 
It can be described by a dimension-six operator~\cite{Wolfenstein:1977ue}
\begin{equation}
\mathcal{L}_{\mathrm{NC-NSI}} \;=\;
-2\sqrt{2}G_F 
\varepsilon_{\alpha\beta}^{fC}
\bigl(\overline{\nu_\alpha}\gamma^\mu P_L \nu_\beta\bigr)
\bigl(\overline{f}\gamma_\mu P_C f\bigr)
\;,
\label{H_NC-NSI}
\end{equation}
where $\alpha, \beta = e,\mu,\tau$ refer to the 
neutrino flavor,  $f = e,u,d$ indicates the matter 
fermions, superscript $C=L, R$ denotes the chirality of the 
$ff$ current, and $\varepsilon_{\alpha\beta}^{fC}$ are the strengths 
of the NSI. The hermiticity of the interaction requires, $\varepsilon_{\beta\alpha}^{fC} \;=\; (\varepsilon_{\alpha\beta}^{fC})^*$.
For neutrino propagation in Earth matter, the relevant combinations are
\begin{equation}
\varepsilon_{\alpha\beta}
\;\equiv\; 
\sum_{f=e,u,d}
\varepsilon_{\alpha\beta}^{f}
\dfrac{N_f}{N_e}
\;\equiv\;
\sum_{f=e,u,d}
\left(
\varepsilon_{\alpha\beta}^{fL}+
\varepsilon_{\alpha\beta}^{fR}
\right)\dfrac{N_f}{N_e}
\;,
\label{epsilondef}
\end{equation}
where $N_f$ is the number density of fermion $f$.
For the Earth, we can assume neutral and isoscalar matter, implying  $N_n \simeq N_p = N_e$, 
in which case $N_u \simeq N_d \simeq 3N_e$. 
Therefore, $\varepsilon_{\alpha\beta}\, \simeq\,
\varepsilon_{\alpha\beta}^{e}
+3\,\varepsilon_{\alpha\beta}^{u}
+3\,\varepsilon_{\alpha\beta}^{d}$.
In presence of NSI, the effective Hamiltonian for neutrino propagation gets modified
in matter, which in turn changes the probability of flavor conversion. For more details please see~\citep{Capozzi:2019iqn}.

In this work, we only focus on the non-diagonal NSIs $\varepsilon_{e\mu}$
and $\varepsilon_{e\tau}$ along with their associated CP-phases $\phi_{e\mu}$ and $\phi_{e\tau}$ respectively.
We recall that the current upper bounds (at 90\% C.L.) on the two NSI under consideration are: $|\varepsilon_{e\mu}| \lesssim 0.12$
and $|\varepsilon_{e\tau}| \lesssim 0.36$ as reported in the review~\cite{Farzan:2017xzy}, which refers to the global 
analysis~\cite{Gonzalez-Garcia:2013usa}. These limits are basically corroborated by the more recent analysis~\cite{Esteban:2019lfo}~(for more details please refer to~\cite{Capozzi:2019iqn}). It is worth to mention that we do not focus on $|\epsilon _{\mu \tau }|$ as it has very strong upper bound $|\epsilon _{\mu \tau }| < 8.0 \times 10^{-3}$~\cite {Aartsen:2017xtt}.
%
Let us now consider the appearance probability relevant for the two LBL experiments T2K and NO$\nu$A under consideration.
In the presence of NSI, the appearance probability can be approximately written as a sum of three terms~\cite{Kikuchi:2008vq}, $P_{\mu e}  \simeq  P_{\rm{0}} + P_{\rm {1}}+   P_{\rm {2}}$. 
The first two terms correspond to the standard 3-flavor probability while the
third one arises due to NSI. Now considering the mixing angle $\theta_{13}$,
the parameter $v (\equiv\, 2V_{CC}E/\Delta m_{31}^2)$ and the coupling $|\varepsilon|$ are of the same order of magnitude $\mathcal{O}(\epsilon)$ and small ($\sim 0.2$)
\footnote{Here we assume that the numerical analysis presented here points toward
best fit values in the range $|\varepsilon| \sim 0.1-0.4$.},
while $\alpha \equiv \Delta m^2_{21}/ \Delta m^2_{31} = \pm 0.03$ is
$\mathcal{O}(\epsilon^2)$, we can expand the above three terms as~\cite{Liao:2016hsa}, 
\begin{eqnarray}
\label{eq:P0}
 & P_{\rm {0}} &\simeq\,  4 s_{13}^2 s^2_{23}  f^2\,,\\
\label{eq:P1}
 & P_{\rm {1}} & \simeq\,   8 s_{13} s_{12} c_{12} s_{23} c_{23} \alpha f g \cos({\Delta + \delta})\,,\\
 \label{eq:P2}
 & P_{\rm {2}} & \simeq\,  8 s_{13} s_{23} v |\varepsilon|   
 [a f^2 \cos(\delta + \phi) + b f g\cos(\Delta + \delta + \phi)]\,,
\end{eqnarray}
where $\Delta \equiv  \Delta m^2_{31}L/4E$ is the atmospheric oscillating frequency, $V_{CC}=\sqrt{2}G_FN_e$ is the charged-current matter potential, 
$L$ being the baseline and $E$ the neutrino energy.  For compactness, we have used the notation
($s_{ij} \equiv \sin \theta_{ij} $, $c_{ij} \equiv \cos \theta_{ij}$), and following~\cite{Barger:2001yr},
we have introduced 
\begin{eqnarray}
\label{eq:S}
&f \equiv & \frac{\sin [(1-v) \Delta]}{1-v}\,, \qquad  g \equiv \frac{\sin v\Delta}{v},\,\\
%
& a = & s^2_{23}, \quad b = c^2_{23}, \quad {\mathrm {if}} \quad \varepsilon = |\varepsilon_{e\mu}|e^{i{\phi_{e\mu}}}\,,\\
& a = & s_{23}c_{23}, \quad b = -s_{23} c_{23}, \quad {\mathrm {if}} \quad \varepsilon = |\varepsilon_{e\tau}|e^{i{\phi_{e\tau}}}\,.
\end{eqnarray}
In the expressions given in Eqs.~(\ref{eq:P0})-(\ref{eq:P2}) for $P_0$, $P_1$ and $P_2$, 
the sign of $\Delta$, $\alpha$ and $v$ is positive (negative) for NO (IO).  
For antineutrinos, the signs in front of the CP-phases and the matter parameter $v$ are flipped in Eqs.~(\ref{eq:P0})-(\ref{eq:P2}).
%
%
\begin{center}
\begin{figure}[h!]
\vspace*{-0.0cm}
\hspace*{2cm}
\includegraphics[height=5.3cm,width=5.3cm]{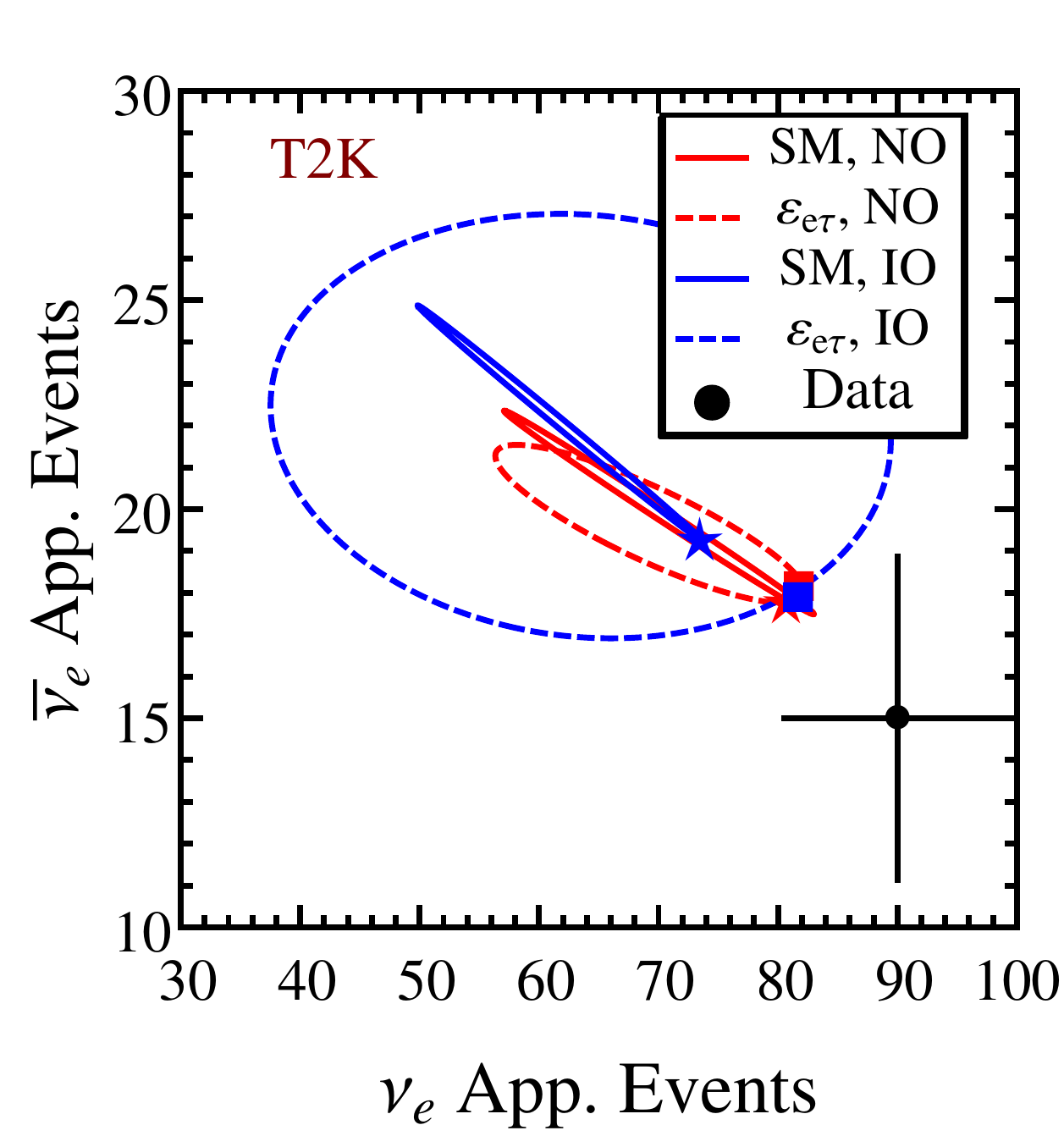}
\hspace*{0.2cm}
\includegraphics[height=5.3cm,width=5.3cm]{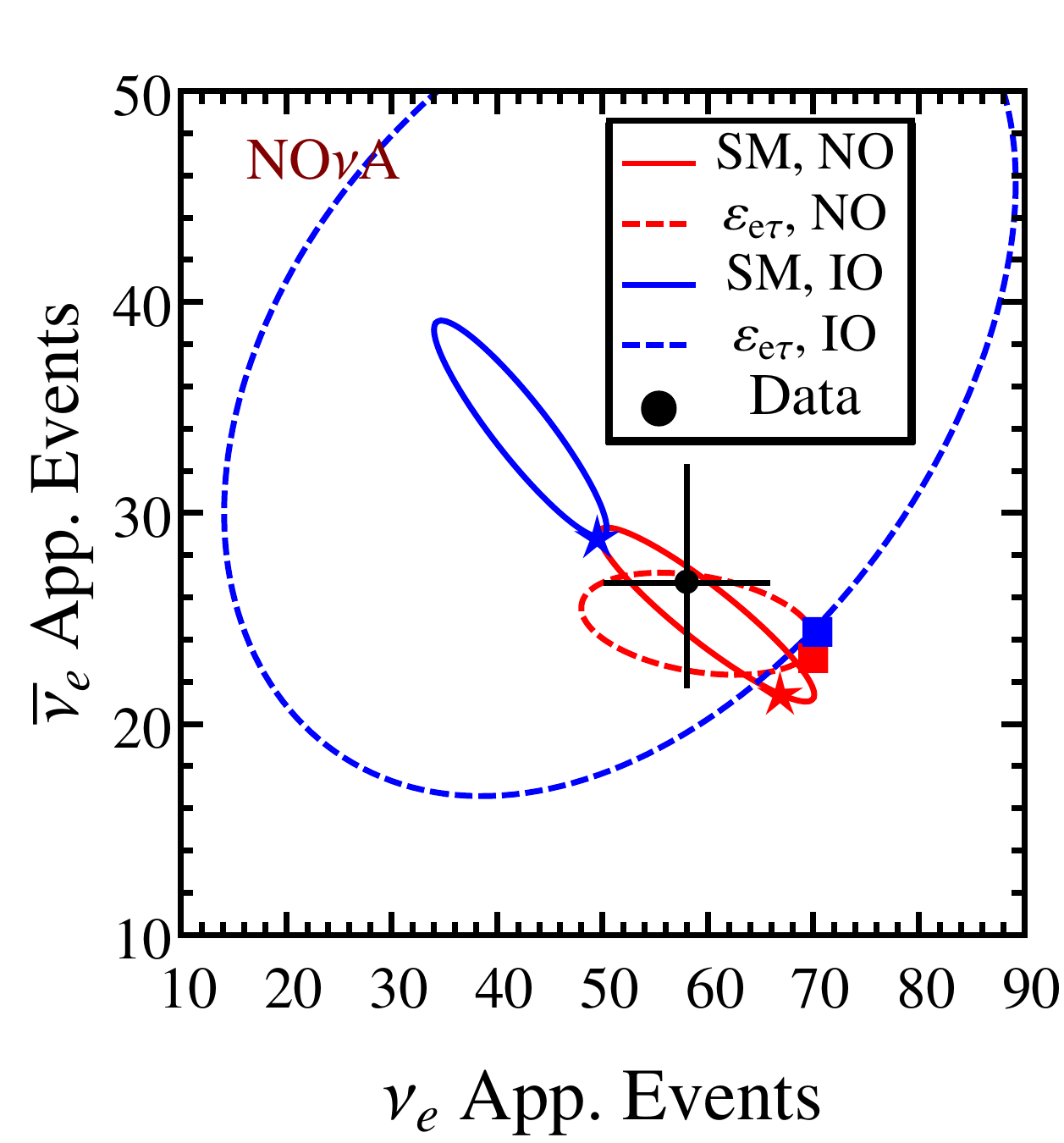}
\vspace*{-0.3cm}
\caption{Bievents plot for the T2K (left panel) and NO$\nu$A (right panel). 
$\delta$ is the running parameter on each ellipse in the range $[0,2\pi]$. The red (blue) star represents the best fit points for SM, NO (IO) determined by fitting the {\em combination} of the two experiments, and similarly the squares correspond to SM+NSI case. This figure has been taken from~\citep{Capozzi:2019iqn}.}
\label{Bievent_1}
\end{figure} 
\end{center}
%
\section{Discussion at the level of the bievents plots}  
\vspace*{-.4cm}
\noindent Before we start the discussion, it is important to mention that for all the analyses presented here we have used the current data of T2K and NO$\nu$A extracted from~\cite{T2K_talk} and~\cite{Acero:2019ksn}. In all our numerical simulations we have fixed the solar parameters $\theta_{12}$ and $\Delta m^2_{21}$ at the best fit point of the global analysis~\cite{Capozzi:2018ubv},
while the reactor angle $\theta_{13}$ has been marginalized away with a strong 3.7\% 1$\sigma$ prior on $\sin^2\theta_{13} = 0.0216$. For more details about the data sets and analysis procedures please see~\citep{Capozzi:2019iqn}. Fig.~\ref{Bievent_1} represents the bievent plots for the two experiments T2K and NO$\nu$A in the context of the current data,  where $x\, (y)$ axis corresponds to neutrino (antineutrino) appearance events. These plots are particularly very useful as they give a rough estimate of the situation of each experiment and clearly express the differences between the different experiments. Left (right) panel shows the bievent ellipses for T2K (NO$\nu$A). The continuous curves correspond to the SM framework and dashed curves correspond to the presence of SM+NSI, particularly $\varepsilon_{e\tau}$. SM (SM + NSI) ellipses have been generated with the fixed (those not marginalized) and the common best fit values of the oscillation parameters obtained from the combined analysis of the two experiments in SM (SM + NSI) framework where the standard CP-phase $\delta$ is the running parameter in the range $[0, 2\pi]$. The best fit values obtained from this analysis can be easily read from Fig.~\ref{fig:fit}. The best fit value for $\Delta m_{31}^2$ (which is not shown in Fig.~\ref{fig:fit}) obtained here in case SM is $2.5\times 10^{-3}$ ($2.46\times 10^{-3}$) $eV^2$ for NO (IO). Similarly in case of SM + NSI ($\varepsilon_{e\tau}$) it is $2.49\times 10^{-3}$ ($2.46\times 10^{-3}$) $eV^2$ for NO (IO).
The black dot points indicate the experimentally observed data for each experiment.  
Let us first comment about the SM results. 
One can see that in case of T2K, the NO best fit (red star) is closer to the data point with respect to the IO best fit (blue star), whereas in NO$\nu$A there is basically no particular preference of any of the two mass orderings. As a result, an overall global preference of NO is obtained at 2.4$\sigma$ as can be seen in Fig.~\ref{fig:fit}. Now in the presence of NSI, the NO and IO
best fit points (red and blue squares) for each experiment basically lie at the same distance from the experimental data points
in contrast to the SM case. We find $\chi^2_{\mathrm{SM+NSI},\, \mathrm{NO}} - \chi^2_{\mathrm{SM+NSI},\, \mathrm {IO}} \simeq 0.5$, corresponding to 0.7$\sigma$ preference of IO over NO (see also Fig.~\ref{fig:fit}).
Therefore, the indication in favor of NO found in the standard case gets completely lost in the presence of NSI. For more details please refer to~\citep{Capozzi:2019iqn}.
%
\begin{figure}[t!]
\vspace*{-0.0cm}
\hspace*{2cm}
\includegraphics[height=5.3cm,width=5.3cm]{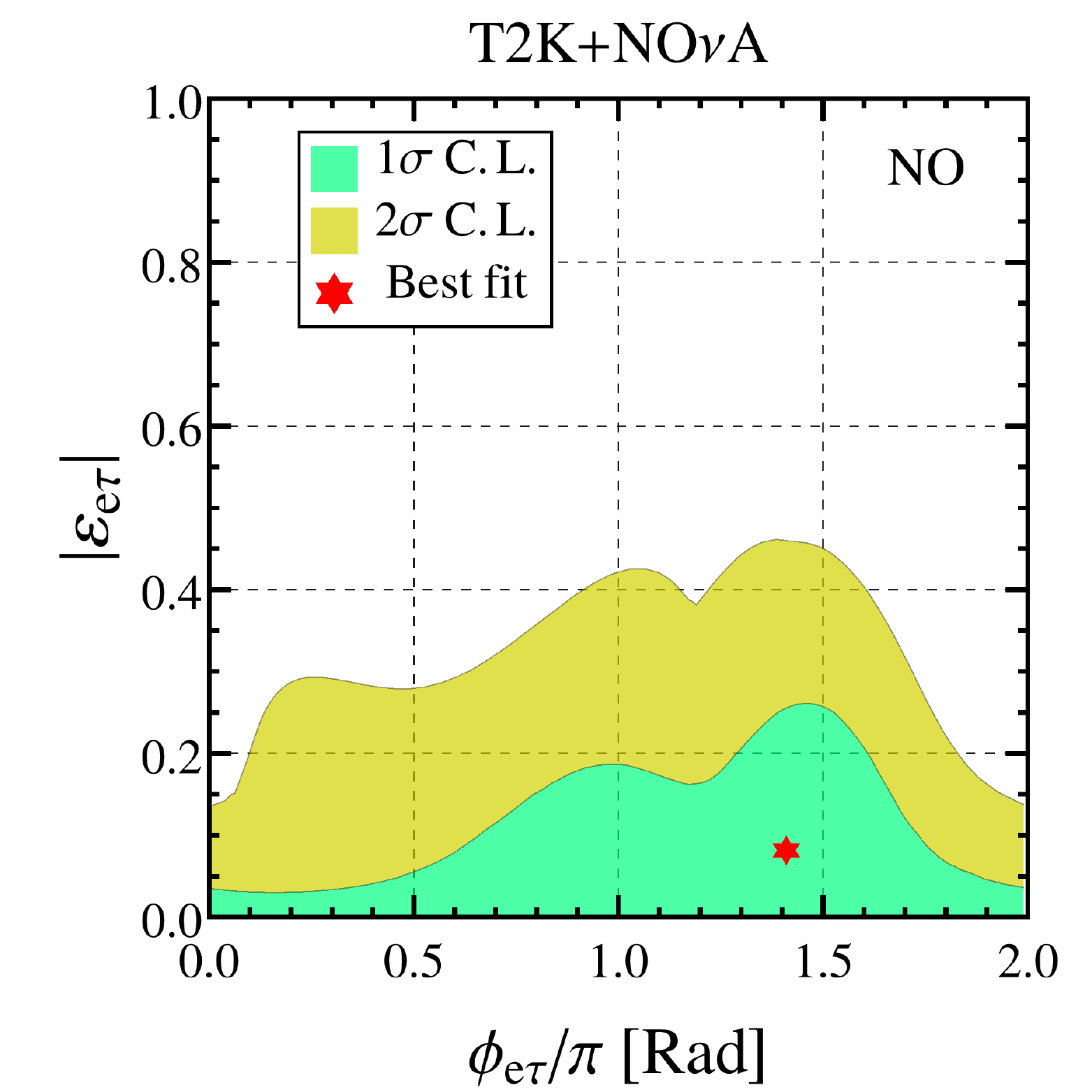}
\hspace*{0.2cm}
\includegraphics[height=5.3cm,width=5.3cm]{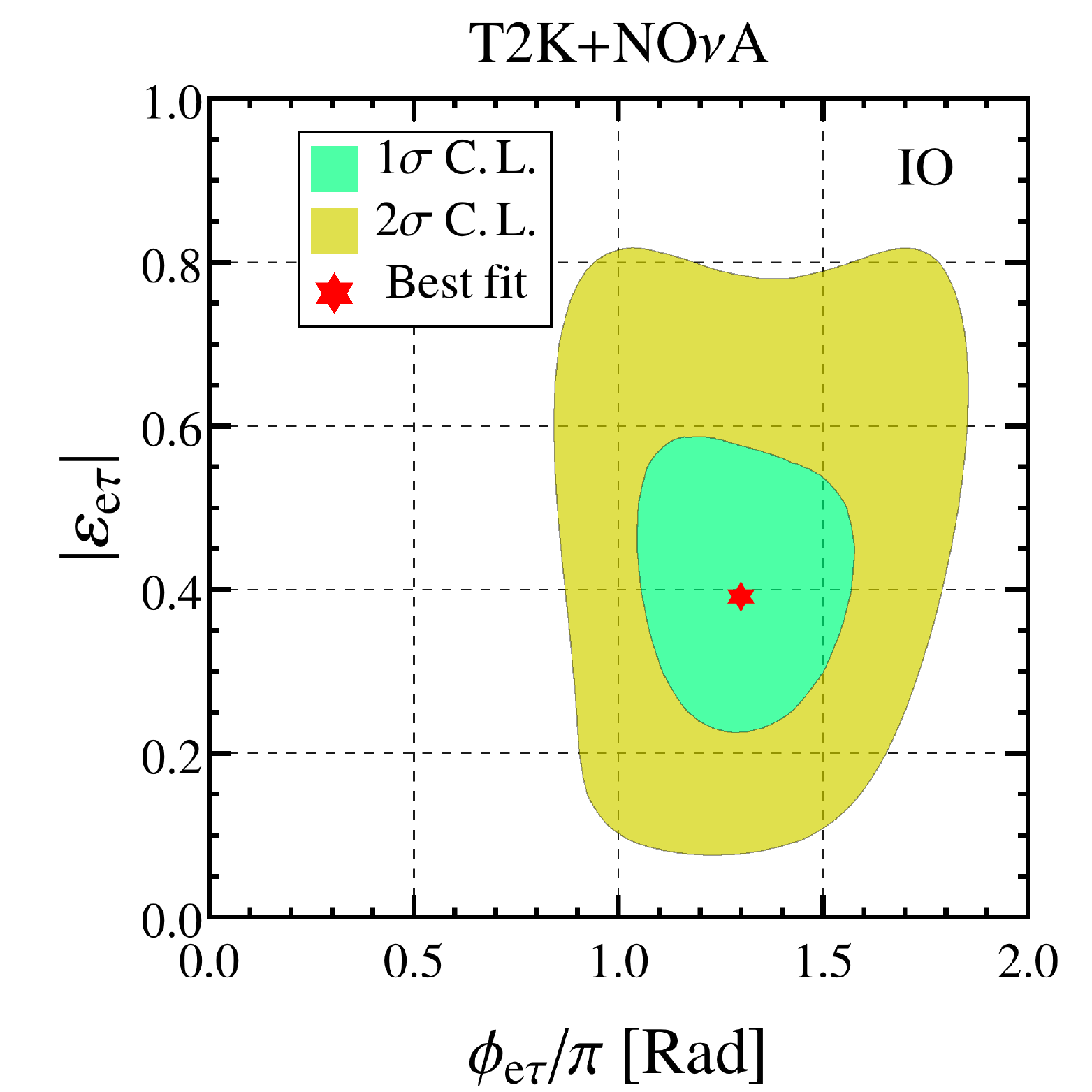}
\vspace*{-0.3cm}
\caption{Allowed regions of the parameters $\left[\phi_{e\tau},\,|\varepsilon_{e\tau}|\right]$ determined by the combination of T2K and NO$\nu$A for NO (left panel) and IO (right panel). The contours are drawn at the 1$\sigma$ and 2$\sigma$ level for 1 d.o.f.. This figure has been taken from~\citep{Capozzi:2019iqn}.}
\label{fig:regions}
\end{figure} 
%
\section{Numerical Results}
\vspace*{-.4cm}
\noindent Fig.~\ref{fig:regions} shows the allowed regions determined by the combined analysis of T2K and NO$\nu$A, in the plane spanned by $|\epsilon_{e\tau}|$ and $\phi_{e\tau}$. Left (right) panel corresponds to NO (IO). The red star in each panel denotes the best fit value. 
We have marginalized away the CP-phase $\delta$, the mixing angles $\theta_{13}$ (with prior) and $\theta_{23}$, and
the squared-mass $\Delta m^2_{31}$. It is evident from the left panel that in case of NO, there is only $0.7\sigma$ level preference of non-zero value of $|\epsilon_{e\tau}|$ with best fit value approximately $0.09$. Now in case of IO, we can see from the right panel there is a $2.5\sigma$ level strong preference of non-zero value of $|\epsilon_{e\tau}|$ with best fit value approximately $0.39$. The associated CP-phase $\phi_{e\tau}$ has the best fit value $1.42\pi$ ($1.30\pi$) for NO (IO).

\begin{figure}[t!]
\vspace*{-0.0cm}
\hspace*{-0.2cm}
\includegraphics[height=4.3cm,width=3.7cm]{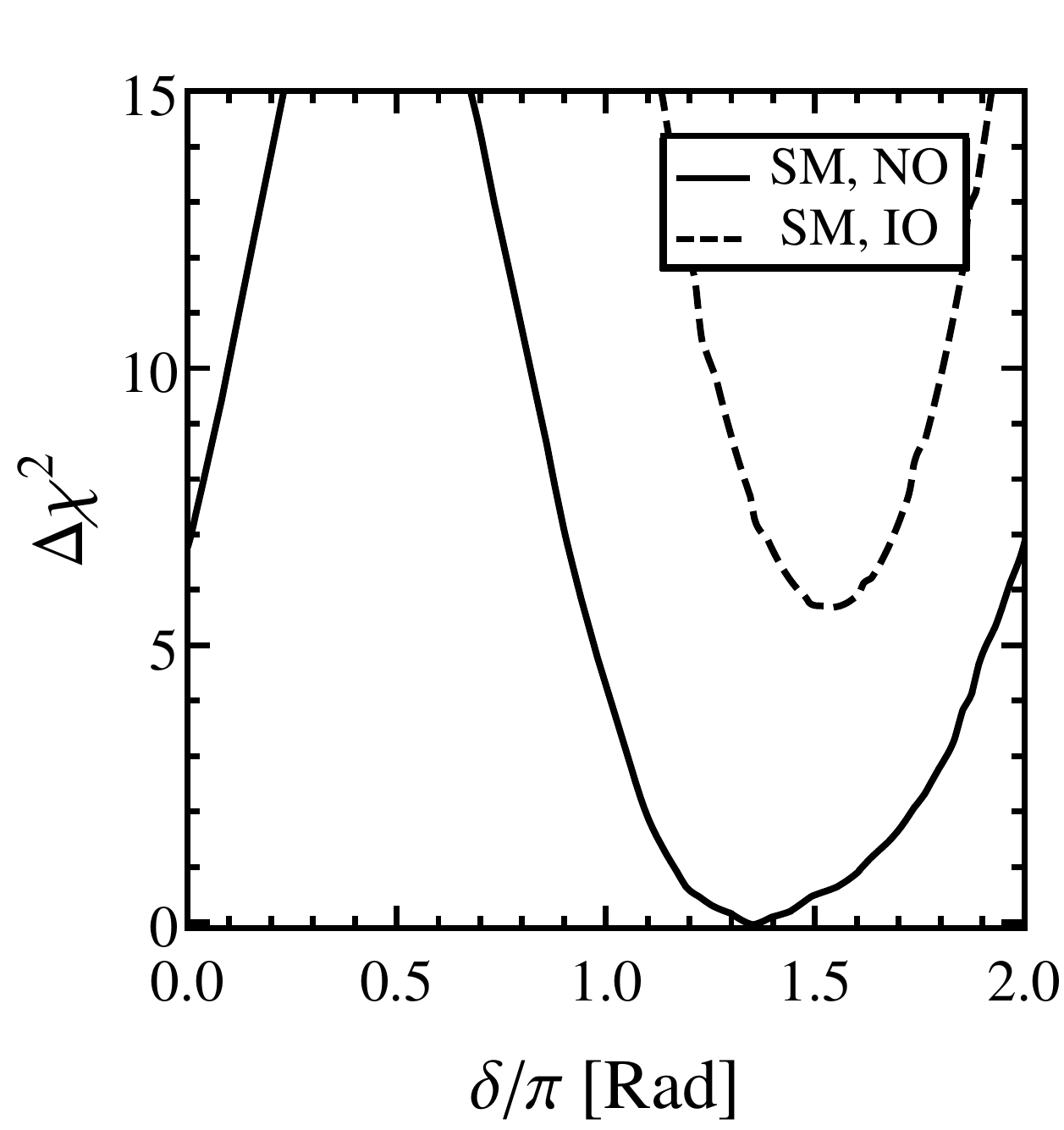}
\includegraphics[height=4.3cm,width=3.7cm]{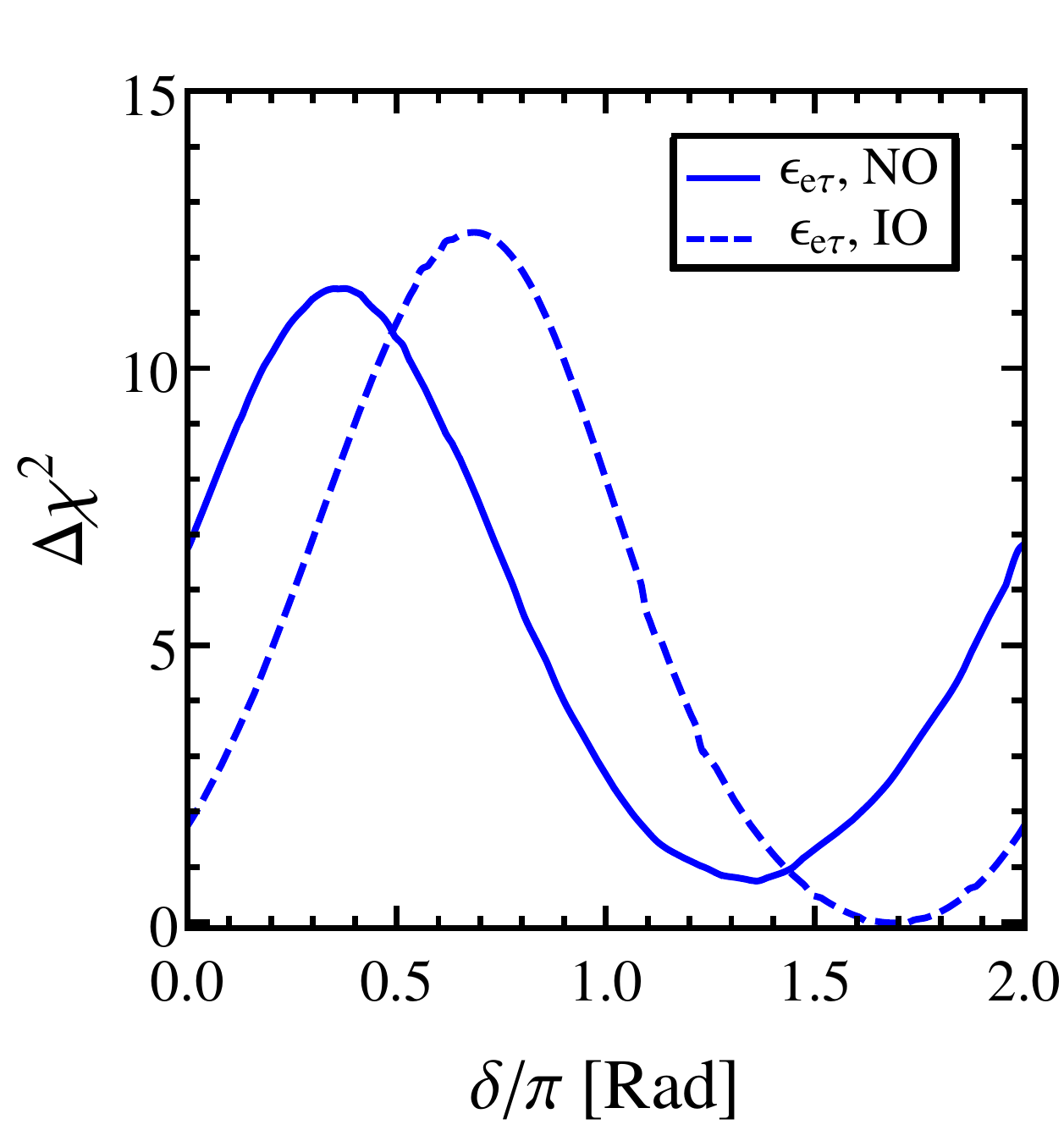}
\includegraphics[height=4.3cm,width=3.7cm]{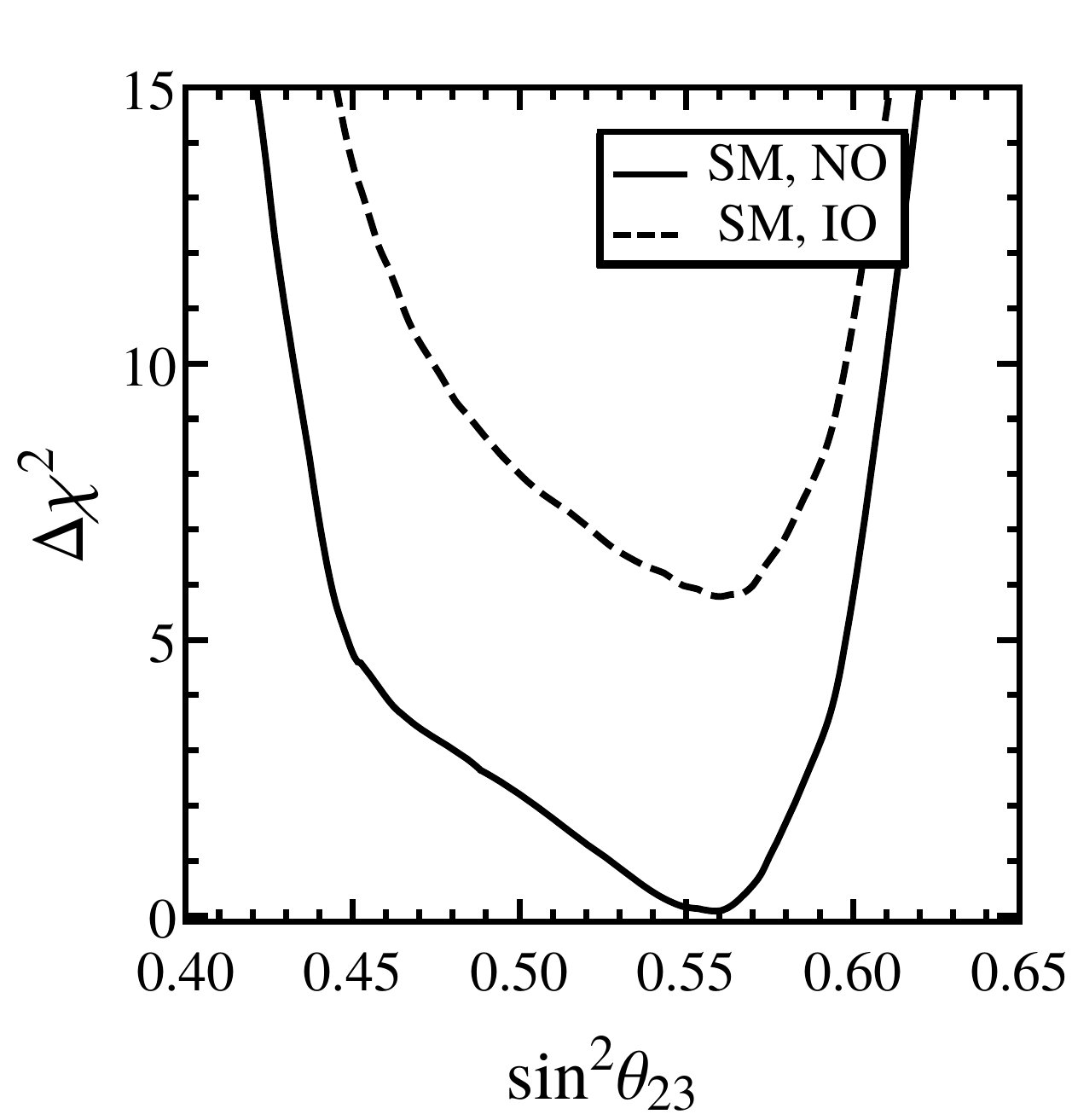}
\includegraphics[height=4.3cm,width=3.7cm]{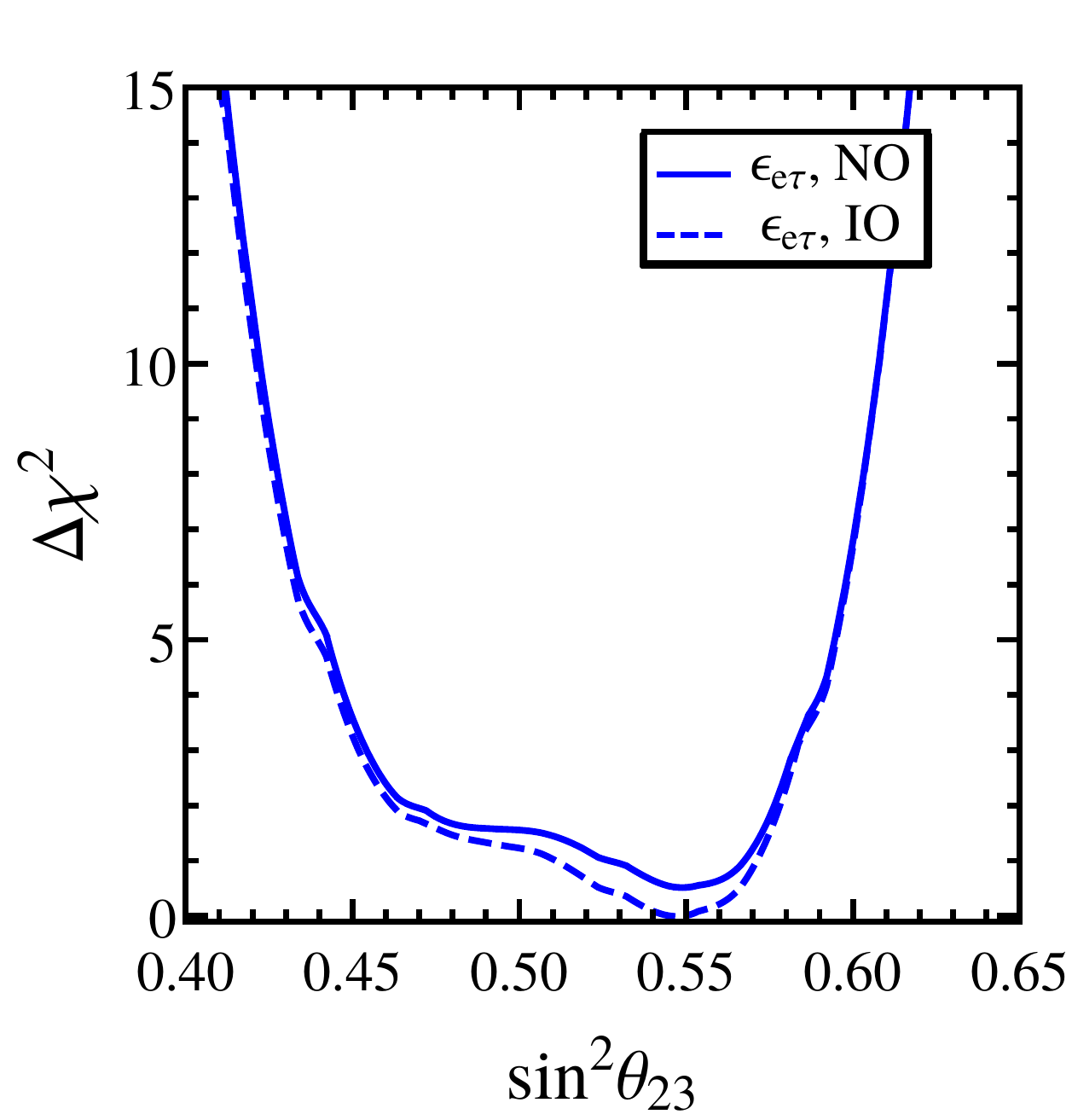}
\vspace*{-0.2cm}
\caption{One dimensional $\Delta\chi^2$ projections of $\delta$ and $\theta_{23}$ for SM as well as SM+NSI framework, determined 
by the combination of T2K and NO$\nu$A. This figure has been taken from~\citep{Capozzi:2019iqn}.}
\label{fig:fit}
\end{figure}

In Fig.~\ref{fig:fit}, we show the one-dimentional projections on two oscillation parameters $\delta$ and $\theta_{23}$ in the SM and SM+NSI framework. The allowed regions shown here are determined by the combined analysis of T2K and NO$\nu$A. Solid (dashed) lines correspond to NO (IO). The undisplayed parameters in each plot have been marginalized away. One can observe that in the SM case there is a 
preference for NO at the $\sim 2.4\sigma$ level. However this preference gets completely washed out and there is no preference of any ordering in the presence of NSI. It is worth to mention that the reconstruction of $\delta$ deteriorates
in the presence of NSI and similarly the preferences for a non-maximal mixing and that for the higher octant of $\theta_{23}$ with respect to the lower octant found in the SM sensibly decrease, which can be attributed to the fact that in presence of NSI the number of degrees of freedom is higher than the SM case. Let us now make some comments regarding the NSI involving the $e-\mu$ sector. We have found that in this case NO is preferred over IO at
the $2.5\sigma$ level similar to the SM case. Therefore, the ambiguity of the indication in favor of the NO appears
only when we consider the non-standard interactions in the $e-\tau$ sector. For more details please see~\citep{Capozzi:2019iqn}.
%
%
%
%
\section{Conclusions}
\vspace*{-.4cm}
\noindent In this work we have expounded the current data of T2K and NO$\nu$A in the presence of NSI involving the $e-\tau$ ($\varepsilon_{e\tau}$) and $e-\mu$ ($\varepsilon_{e\mu}$) sectors respectively. We found that the indication in favor of NO in SM gets completely vanished when one turns on the NSI with coupling strength ($\varepsilon_{e\tau}$).  It would be interesting to make the same analysis combining  
NO$\nu$A and T2K data with that of the existing atmospheric neutrino data. We also hope that the future high-statistics LBL experiments like DUNE and the
atmospheric neutrino facilities might help to resolve this issue.

\acknowledgments
\vspace{-.25cm} 
\noindent S.S.C. would like to thank the organizers of ``40th International Conference on High Energy physics - ICHEP2020'' for giving an opportunity to present this work.

\bibliographystyle{JHEP}
\bibliography{NSI-References}
%

\end{document}